\documentclass[showpacs,aps,prb,twocolumn,groupedaddress,amsmath,amssymb]{revtex4}

\usepackage{graphicx}
\usepackage{color}

\newcommand{\tr}{\ensuremath{\operatorname{tr}}}

\newcommand{\ev}[1]{\ensuremath{\langle #1 \rangle}}

\newcommand{\abs}[1]{\ensuremath{|#1|}}

\begin{document}
	\title{Adiabaticity in semiclassical nanoelectromechanical systems}
	
	\author{A. Metelmann}
	\email[]{metelmann@itp.tu-berlin.de}
	\author{T. Brandes}
	\affiliation{Institut f\"ur Theoretische Physik, TU Berlin, Hardenbergstr. 36, D-10623 Berlin, Germany}
	\date{\today}
	
\begin{abstract}
We compare the semiclassical description of NEMS within and beyond the adiabatic approximation.
We consider a NEMS model which contains a single phonon (oscillator) mode linearly coupled to an electronic few-level system in contact with external particle reservoirs (leads). 
Using Feynman-Vernon influence functional theory, we derive a Langevin equation for the oscillator trajectory that is non-perturbative in the system-leads coupling. A stationary electronic current through the system generates nontrivial dynamical behavior of the oscillator, even in the adiabatic regime. The \lq backaction\rq  \ of the oscillator onto the current is studied as well. 
For the two simplest cases of one and two coupled electronic levels, we discuss the differences between the adiabatic and the non-adiabatic regime of the oscillator dynamics.  
\end{abstract}

	
\maketitle

\section{Introduction}
Nanoelectromechanical systems (NEMS) enable the detailed study of the interaction between electrons, tunneling through a nano-scale device, and the degrees of freedom of a mechanical system. The electronic current affects the mechanical system and vice versa. The dimensions of systems used in recent experiments range down to scales, where the observation of fundamental quantum behavior for a comparatively macroscopic object is possible \cite{OConnelletal2010, Teufel2011}.

The influence of strong electron-phonon coupling in molecules or suspended quantum dots yields highly interesting effects \cite{Lassange2009, Leturcq2009, Sampmaz2006}, like the Franck-Condon blockade, where the influence of the mechanical system suppresses the electronic current \cite{Leturcq2009, Sampmaz2006}, or switching in molecular junctions \cite{Blum2005}. 

In non-equilibrium, a standard approach to solve the dynamics of NEMS is to do perturbation theory in a tunnel Hamiltonian, which has been successfully done up to the co-tunneling regime \cite{Wegewijs2008}. This is a well-explored path, where the dynamics is described by master equations or generalizations thereof in Liouville space. Many interesting physical results can be obtained via this approach, e.g. avalanche-type molecular transport \cite{KochAll} or laser-like instabilities \cite{Rodrigues2007}.
These methods produce good results in the range of high bias, where non-Markovian effects can be neglected. To gain access to the small bias regime,
one has to work perturbatively in the system-oscillator instead of the system-leads coupling \cite{Mitra04}. Alternatively, if the oscillator is treated in a semiclassical regime, Feynman-Vernon influence functional techniques are suitable \cite{Feynman1963, Mozyrsky2004,Mozyrsky2006, PhysRevLett.72.2919, PhysRevB.52.6042}. 

In a previous work, we combined a semiclassical analysis with an adiabatic approach \cite{Hussein2010}, where we assumed the oscillator movement to be slow compared to the electrons which are jumping through the system. The condition $\Gamma \gg \omega_{0}$ then followed from this approximation, where $\Gamma$ denotes the tunneling rate of the electrons and $\omega_{0}$ is the oscillator frequency.
The most interesting results, such as negative damping and limit cycles, were achieved in the regime where the oscillator and the electrons act on the same time scale. But in the latter regime, the adiabatic approach was at the limits of its validity.

In recent publications different approaches, e.g. based on scattering theory \cite{Bennett2010, Bode2011}, were used to go beyond the adiabatic approximation or to verify the range of validity for this approach \cite{Nocera2011,Piovano2011}. In this paper, we go one step further and present numerical results for a completely non-adiabatic approach. We apply our method to two simple NEMS models, the single and the two-level system.
We work in the semiclassical regime where an expansion around the classical path is performed. The advantage of this method is that we are non-perturbative in the system-leads coupling, because the exact electronic solutions are included.
In the non-adiabatic approach, we can treat the oscillator and the electrons on the same time scale without further constrains. Therefore, we modify our adiabatic path-integral approach, obtaining an explicit time-dependent perturbation. As a consequence, we have to calculate system quantities numerically in a full time dependent manner. This allows us to critically assess the validity of the adiabatic results. 
By comparing the outcomes of the adiabatic and the non-adiabatic approach we find a qualitative accordance. The same features arise in both approaches and the stationary results predominantly coincide, but there is no quantitative accordance in the time-dependent regimes. The differences increase together with the complexity of the focused NEMS model.

This paper is organized as follows. In Sec. \ref{sec.:model} we introduce the general model, followed by the derivation of the stochastic equation of motion for the oscillator dynamics.
In Sec. \ref{sec.:singlelevel} we present the results for a single resonant level system, comparing the adiabatic and the non-adiabatic regime. In addition to the phase space trajectories, we calculate the resulting electronic current through the system. In Sec. \ref{sec.:twolevel} we consider a two-level system, where the dynamical behavior of the oscillator exhibits non-trivial effects. Thereby we show that the  oscillations of the mechanical subsystem lead to an oscillating current with fixed frequency.

\section{Model} \label{sec.:model}
Our total Hamiltonian is a sum of an electronic system $\mathcal H_{\rm e}$, a single oscillator with a spatial degree of freedom $\hat q$ in a harmonic potential $\mathcal H_{\rm osc}$ and a linear coupling between the oscillator and the electronic system
    \begin{align}
	\mathcal H &= \mathcal H_{\rm e} + \mathcal H_{\rm osc} - \hat F \hat q, \label{eq.:H_gen}
    \end{align}
in which $\hat F$ denotes an electronic force operator. The electronic system itself consists of a few electronic levels which are connected to two macroscopic leads. The latter are considered as two Fermi seas with chemical potential $\mu_{\alpha \in{\rm{L,R}}}$ and temperature $T$. Furthermore, the electronic part provides a non-equilibrium environment for the system. In our formalism we work non-perturbatively in the system-leads coupling, assuming arbitrary coupling and a finite bias regime without constrains.

The single oscillator with momentum $\hat p$, position $\hat q$ and mass $m$ is described by a parabolic potential 
    \begin{align} \label{eq.:H_osc}
	  \mathcal H_{\rm osc}=\frac 1{2m} \hat p^2 +\frac12 m\omega_0^2 \hat q^2,
    \end{align}
whereby $\omega_0$ equals the oscillator frequency.
The oscillator potential will be modified by the electronic environment and will exhibit multi-stabilities due to the electronic forces\cite{Hussein2010}.  
In this paper, the reduced Planck constant is set to unity ($\hbar = 1$).
	
\subsection{Influence functional}
We want to focus on the oscillator's dynamics, which is described by the reduced density matrix $\rho_{\rm osc}(t)$, obtained from the total density matrix $\chi(t)$ by tracing out the bath degrees of freedom, $\rho_{\rm osc}(t) = \tr_{\rm B} \chi(t) $.
The propagation of the reduced density matrix in time can be written as a double path integral over $q_{t}$ (forward) and $q'_{t}$ (backward) weighted by the Feyman-Vernon influence functional \cite{Weiss2008, Schulman2005}
    \begin{equation} \label{eq.FeymannInfluence}
	   \mathcal{F}\left[ q_{t} ; q'_{t} \right] = \tr_{\rm B} ( U^{\dag}\left[ q'_{t} \right] U\left[ q_{t} \right]), 
    \end{equation}
containing the time evolution operator
    \begin{equation}
         U\left[ q_{t} \right]  = T e^{- i \int\limits_{0}^{t} dt' \ \left[\mathcal H_{e} - \hat{F} q _{t'} \right] }.
    \end{equation}
The time-ordering operator $T$ arranges operators with later times to the left.

We want to perform an expansion around the classical path. Therefore we transform to center-of-mass and relative path variables 
    \begin{equation}
	  q_{t} = x_{t} + \frac{1}{2} y_{t}, \hspace{1cm} q'_{t} = x_{t} - \frac{1}{2} y_{t},
    \end{equation}
where the variable $y_{t}$ can be interpreted as the quantum fluctuations around the classical path.
We introduce an interaction picture 
    \begin{eqnarray} \label{eq.TimeEvolutionOperator}
	  U\left[ q_{t} \right]& = & T e^{- i \int\limits_{0}^{t} dt' \ \left[\hat{H}_{e} - \hat{F} x_{t'} -  \frac{1}{2} \hat{F} y_{t'}\right] }
                              \nonumber \\
                          & = & U\left[ x_{t} \right] \widetilde{U}\left[ y_{t} \right] ; \hspace{0.3cm} \widetilde{U}\left[ y_{t} \right] 
                            = T e^{i \int\limits_{0}^{t} dt' \ \frac{1}{2} \widetilde{F}(t') y_{t'}},
    \end{eqnarray}       
where the term with the off-diagonal path $y_t$ is regarded as a perturbation and $\widetilde{F}(t) = U^{\dag}\left[ x_{t} \right] \ \hat{F} \ U\left[ x_{t} \right]$.
Inserting ~Eq.~\eqref{eq.TimeEvolutionOperator} into the influence functional, ~Eq.~\eqref{eq.FeymannInfluence}, leads to
    \begin {eqnarray} \label{eq.FeymannInfluence_2}
         \mathcal{F}\left[ q_{t} ; q'_{t} \right] 
                    & = & \tr_{\rm B} ( \widetilde{U}^{\dag}\left[ - y_{t} \right] U^{\dag}\left[ x_{t} \right]
                                                  U\left[ x_{t} \right] \widetilde{U}\left[ y_{t} \right]) \nonumber \\ 
                    & = & \tr_{\rm B} ( \widetilde{U}^{\dag}\left[ - y_{t} \right] \widetilde{U}\left[ y_{t} \right]).
    \end{eqnarray}
Expanding this term to second order and performing a cluster expansion \cite{VanKampen2008}, we finally obtain 
    \begin{equation} 
          \mathcal{F}^{pert} \left[ q_{t} ; q'_{t} \right] = e^{- \Phi \left[ x_{t} ; y_{t} \right] },
    \end{equation}
with the influence phase
    \begin{align} \label{eq.:FeymannInfluence_3}
           \Phi \left[ x_{t} ; y_{t} \right] =&
               - i \int\limits_{0}^{t}  dt'  f(t')  y_{t'} +  \int\limits_{0}^{t}  dt' \int\limits_{0}^{t}  ds  C(t',s)  y_{t'}  y_{s},
    \end{align}
and the force correlation function
    \begin{align}
            C(t',s) =&  \tr_{\rm B} \left\lbrace \left( \widetilde{F}(t') - f(t') \right) \left( \widetilde{F}(s) - f(s) \right)\right\rbrace
                    \nonumber \\
                    \equiv &  \langle \delta\widetilde{F}(t') \delta\widetilde{F}(s)\rangle.
    \end{align} 
The force term $f(t) \equiv \langle \widetilde{F}(t) \rangle$ depends on the center of mass path $x_{t}$.
The influence phase,~Eq.~\eqref{eq.:FeymannInfluence_3}, can be regarded as a cumulant generating functional for the force operator correlation functions. The term of quadratic order in $y_t$ describes the Gaussian fluctuations around the classical trajectory which is determined self-consistently in our approach. Higher order terms in $y_t$ (corresponding to non-Gaussian noise) describe higher quantum fluctuations which are neglected here.

\subsection{Langevin equation}
To second order in $y_t$ the double path integral for the reduced density matrix describes a classical stochastic process for the diagonal path $x_t$ that is defined by a (non-adiabatic) Langevin equation
    \begin{equation} \label{eq.LangevinGeneric}
           m  \mbox{\"{x}}_{t} + V'_{\rm osc} (x_{t}) - f \left[ x_{t} \right] = \xi_{t} 
    \end{equation}
with $V'_{\rm osc} (x_{t}) = m \omega_0^2 x_t$ and a Gaussian stochastic force $\xi_{t}$ that has a correlation function $\langle \xi_{t'} \xi_{s}\rangle = C(t',s)$. ~Eq.~\eqref{eq.LangevinGeneric} is the starting point for our non-adiabatic calculations. Note that the force $f \left[ x_{t} \right]$ is a complicated functional that contains the full time-dependence of the position operator $x_t$.

In the adiabatic approximation, a Taylor expansion for the center of mass variable is performed  ($x_t \approx x_0 + t * \dot{x}_0$), leading to an interaction picture with respect to $\mathcal{H}_0 = \mathcal{H}_{\rm e}-\hat F x_0 $ and a perturbation $V[q](t)= -\hat F(t{\dot x}_0+\frac12 y_t)$. Consequently, the expectation value of the force operator in the adiabatic interaction picture can be calculated for fixed $x_0$.
Additionally, due to the second term $t \dot{x}_0$ of the Taylor expansion, an explicit friction term arises in the (adiabatic) Langevin equation,
    \begin{align} \label{eq.:LangevinGenAdiabatic}
	  m  \mbox{\"{x}}_{t} + V'_{\rm osc} (x_{t}) + \dot{x}_t D\left[ x_t \right] - \widetilde f \left[ x_{t} \right] = \xi_{t}. 
    \end{align}
The friction term $D\left[ x_t \right]$ can be interpreted as the first adiabatic correction term. 

In contrast, in the non-adiabatic Eq.~\eqref{eq.LangevinGeneric}, all higher orders of the Taylor expansion are included and the first challenge is to calculate the force term $f \left[ x_{t} \right]$ considering the full time-dependence of $x_t$. In all our calculations presented in this paper we neglect the stochastic fluctuations  ($\xi_t=0$).

\section{Single resonant level} \label{sec.:singlelevel}
    \begin{figure}[t]
		\centering
		\includegraphics[width=\linewidth]{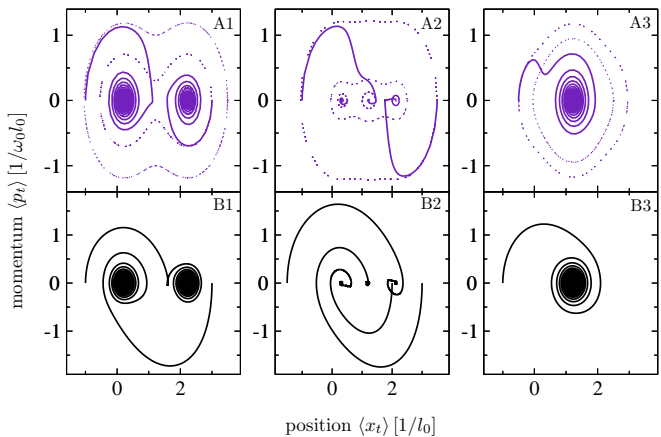}		\caption{\label{fig.:SET_phase_space_comparison}
		Phase space portraits resulting from the (non-) adiabatic approach in units of $\omega_0$ and $l_0$ with the parameters $\Gamma = 1.4 \omega_0$, $\varepsilon_{\rm d} = 3.0 \omega_0$ and $g=2.45$ at zero temperature. Row A depicts the adiabatic results, additionally the case without friction is plotted (dotted lines). Row (B) depicts the non-adiabatic results. The bias voltage (symmetric choice) is increased from left to right, explicit values are $V_{\rm bias}/\omega_0=0.5/2.5/5.0$.
		}
    \end{figure}
The Hamiltonian for the single resonant level (Anderson-Holstein model -- AHM) is
    \begin{equation}
       \mathcal H =  \sum_{k \alpha}  \varepsilon_{k \alpha} \hat c_{k \alpha}^{\dag} \hat c_{k \alpha}  
                +  \sum_{k \alpha} \left( V_{k \alpha}  \hat c_{k \alpha}^{\dag} \hat d  
                +  h.c. \right) 
                +  \varepsilon \ \hat d^{\dag} \hat d + \mathcal H_{\rm{osc}},                 
    \end{equation}
with the abbreviation $\varepsilon =  \varepsilon_{d} - \lambda \hat q$, containing the energy of the local level 
$\varepsilon_d$ and the linear coupling to the oscillator. Thereby, $\lambda$ equates the coupling strength. The operators $\hat d/ \hat d^{\dag}$ correspond to the dot, and the lead operators $\hat c_{k \alpha}/ \hat c_{k \alpha}^{\dag}$ annihilate/create an electron in the $\alpha$-lead with energy $\varepsilon_{k \alpha}$ and momentum $k$, whereby $\alpha = \rm L,R$ denote the left/right lead. Transitions of electrons between the dot and the $\alpha$-lead are possible with the amplitude $V_{k \alpha}$.
Here, the force operator becomes $\hat F = \lambda \hat d^{\dag}  \hat d = \lambda \hat n_d $ and the self-consistent equation of motion for the expectation value of the oscillator coordinate $x_t$ (neglecting stochastic fluctuations) reads
    \begin{eqnarray} \label{eq.LangevinSingle}
           m  \mbox{\"{x}}_{t} &+& V'_{\rm osc} (x_{t}) - \lambda N \left[ x_{t}\right] = 0 \nonumber \\
            N \left[ x_{t}\right] &\equiv& \tr_{\rm B} \left( U^{\dag} \left[x_{t}\right] \hat{n}_d U \left[x_{t}\right] \right).
    \end{eqnarray}
The occupation number of the local level $ N \left[ x_{t}\right]$ is calculated with the lesser Green function
    \begin{align}
            N \left[ x_{t}\right] =& \langle \tilde d^{\dag}(t) \tilde d(t) \rangle  = - i G^{<} (t,t),
    \end{align}
which is obtained from  the Keldysh equation
    \begin{equation}
          G^{<} (t,t) = \int dt_{1} \int dt_{2} \ G^{r}(t,t_{1}) \ \Sigma^{<} (t_{1},t_{2}) \ G^{a} (t_{2},t),
    \end{equation}
containing the advanced/retarded Green function
    \begin{equation}
           G^{r,a} (t,t') = \mp i \Theta(\pm t \mp t') \ e^{-i \int\limits_{t'}^{t} dt'' \left[ \varepsilon(t'') \mp i \frac{\Gamma}{2} \right] },
    \end{equation}
with $\varepsilon (t) \equiv \varepsilon_d - \lambda x_t$. Thereby, $\Theta$ designates the Heaviside step function.
In the time dependent case the lesser self energy \cite{Haug2008} reads
    \begin{equation}
          \Sigma^{<}(t_{1},t_{2}) = i \sum_{\alpha \in \rm{L,R}} \int \frac{d\omega}{2\pi} \ 
                                   e^{-i\omega (t_{1} - t_{2})} \ f_{\alpha}(\omega) \ \Gamma^{\alpha},
    \end{equation}
whereby $f_{\alpha}(\omega)$ denotes the Fermi function.
We assume constant tunneling rates $\Gamma_{\alpha} = 2 \pi \sum_k \abs{V_{k\alpha}}^2 \delta(\omega - \varepsilon_{k\alpha}) = \Gamma/2$, i.e. the left and the right tunneling rate are equal with $\Gamma \equiv \Gamma_{\rm L} + \Gamma_{\rm R}$. As a result, for the time dependent occupation we obtain
    \begin{equation}
         N \left[ x_{t}\right] = \sum_{\alpha \in \rm{L,R}} \ \Gamma_{\alpha} \int \frac{d\omega}{2\pi} \ 
                                                       f_{\alpha}(\omega) \ \lvert A (\omega, t)\rvert^{2},
    \end{equation} 
with the spectral function 
    \begin{equation} \label{eq.Spectralfunction}
         A (\omega, t) = - i \ \int\limits_{t_{0}}^{t} dt' \  e^{- i \int\limits_{t'}^{t} dt''
                                 \ \left(\varepsilon (t'') - \omega - i\frac{\Gamma}{2} \right)}.
    \end{equation}
To solve the equation of motion without any further approximation and expansion, we transform Eq.~\eqref{eq.Spectralfunction} into the differential equation
    \begin{equation} \label{eq.spectralDiff}
          \dot{A} (\omega, t) = -i - i \left( \varepsilon_d - \lambda x_t - \omega - i\frac{\Gamma}{2} \right) A (\omega, t).
    \end{equation}
For the numerical integration a trapezoidal rule for discrete functions is applied.
Thus we solve ~Eq.~\eqref{eq.spectralDiff} together with the system 
    \begin{align}
           \dot{x}_t =& \frac{1}{m} p_t \nonumber \\
           \dot{p}_t =& - V'_{\rm{osc}}  
                        + \lambda  \sum_{\alpha \in \rm{L,R}} \frac{\Gamma_{\alpha}}{4\pi} \Delta \omega \nonumber \\  &
                                  \sum_{n=0}^{N-1} \left[ f_{\alpha}(\omega_{n +1}) \ \lvert A (\omega_{n+1}, t)\rvert^{2} +
                                                          f_{\alpha}(\omega_{n}   ) \ \lvert A (\omega_{n}, t  )\rvert^{2} \right] \nonumber \\
     \end{align}
with $\Delta\omega = \lvert \omega_{N}-\omega_{0}\rvert / N $, in which $N$ equals the number of points of the discretization scheme.

Figure \ref{fig.:SET_phase_space_comparison} depicts the results in the oscillator phase space for different bias values. 
All data are obtained for a small tunneling rate $\Gamma = 1.4 \omega_0$, close to the limit of validity of the adiabatic approach which requires $\Gamma \gg  \omega_{0}$. In order to obtain correct physical units we introduce the dimensionless coupling parameter $g = \lambda/ (m \omega_0^2 l_0 )$, whereby  $l_0 \equiv 1/\sqrt{ m \omega_0}$ equals the oscillator length.
Row A shows the adiabatic results. The dotted lines correspond to the adiabatic case without the first adiabatic correction term. Here, the trajectories run about the fixed points of the system. By varying the applied bias, the number of fixed points changes and in the case of high bias only one fixed point survives. Turning on the friction (first adiabatic correction term) leads to the solid line results in the graphs of row A. Here, the centers turn into stable spirals and the trajectories end up in the fixed points. (For further explanation see \cite{Hussein2010}). In row B the non-adiabatic results are depicted. After long times t we observe little variation to the adiabatic result, nevertheless all trajectories end up in the same fixed points. By comparing both approaches the largest differences emerge for small times.    
\begin{figure}[t]
		\centering
		\includegraphics[width=\linewidth]{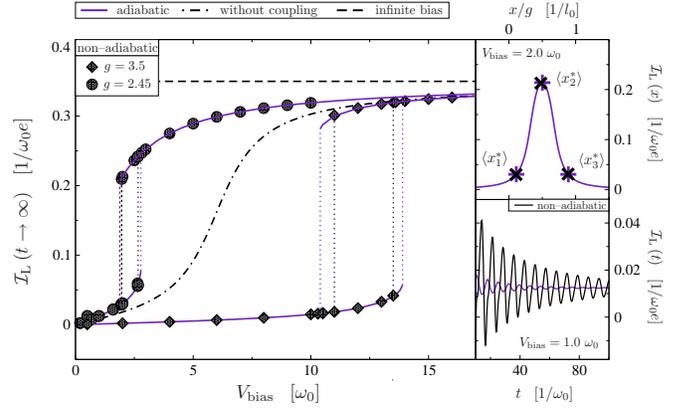}		\caption{\label{fig.:currentSD}
		LEFT: Current for $t \rightarrow \infty$ as a function of $V_{\rm bias}$ for two different coupling parameters $g$ and with $\Gamma=1.4 \omega_0$, $\varepsilon_{\rm d}=3.0 \omega_0$ at zero temperature. The black symbols depict the non-adiabatic and the indigo solid line the adiabatic results. For comparison the infinite bias result (dashed line) and the current without coupling (dashed-dotted line) are plotted. The dotted lines mark the hysteresis like regimes, where two clearly distinct current channels exists.  RIGHT: The upper graph depicts the position dependent current $\mathcal I_{\rm L}(x)$ for $V_{\rm bias}= 2.0 \omega_0$, the three fixed points are marked as crosses. The lower graph shows the left current for small times and $V_{\rm bias}= 1.0 \omega_0$. Remaining parameters are equal to the left graph and $g = 2.45$.}
\end{figure}
This qualitative good accordance can also be observed in the results for the electronic current.
In the non-adiabatic approach, the current is obtained from \cite{Haug2008}
    \begin{align} \label{eq.currentSD}
           \mathcal I_{\alpha} (t) = -e \Gamma_{\alpha} 
                        \left[ N \left[ x_{t}\right] + \int \frac{d\omega}{\pi} f_{\alpha}(\omega) 
                                                                     \mbox{Im} \left[ A (\omega, t)\right] \right].
    \end{align}
The imaginary part of the spectral function is negative and describes the current flowing from the left lead into the dot.
While keeping the oscillator center of mass coordinate $x_t \equiv x$ fixed in Eq.~\eqref{eq.currentSD} when calculating the spectral function, the adiabatic current result is reproduced. Starting from Eq.~\eqref{eq.Spectralfunction}  we obtain
    \begin{align}
          A^{\rm{adiabatic}} (\omega, t) = \frac{ e^{-i\left(\varepsilon_d - \lambda x - \omega - i\frac{\Gamma}{2} \right)(t-t_0) } -1 }
                                                      {\left(\varepsilon_d - \lambda x - \omega - i\frac{\Gamma}{2} \right)}.
    \end{align}
For large times $t-t_0$, the first exponential can be neglected and the spectral function becomes stationary. Therefore the adiabatic current reads (zero temperature)
    \begin{multline}
           \mathcal I_{\rm L} = e \frac{\Gamma}{4 \pi} \bigg[ \arctan \frac{2(\mu_{\rm L} - \varepsilon_d + \lambda x}{\Gamma} \\ - 
                                                                \arctan \frac{2(\mu_{\rm R} - \varepsilon_d + \lambda x}{\Gamma}\bigg] = - \mathcal I_{\rm R}.
    \end{multline}
This is a well known result and for the infinite bias case we obtain $\mathcal I_{L,R}^{\rm{IB}} = \pm e \Gamma/4$ as expected. 
Note, that in the adiabatic case the values for left and right current only differ in their sign. 

In the left graph of Figure \ref{fig.:currentSD} the stationary left current $\mathcal I_{\rm L}(t \rightarrow \infty$) is depicted for increasing bias and for two different coupling parameters $g$. As explained above, the oscillator trajectories end up in fixed points for large times. Hence, the current $\mathcal I_{\rm L} (t\rightarrow \infty)$ becomes stationary and its value corresponds to a single level which is shifted by $ - g x^{\ast}/l_0$, whereby $x^{\ast}$ corresponds to a fixed point.
Because the system owns multiple fixed points, we obtain several current channels depending on the initial condition. The dashed-dotted line in Figure \ref{fig.:currentSD} corresponds to the case without coupling to the oscillator ($g = 0$). 
For small bias, the current is small compared to the infinite bias case (dashed line). There, the effective level $\tilde\varepsilon =\varepsilon_{d} -g x^{\ast}/l_0$ is situated outside the transport window. The latter is also valid for the case without coupling, due to $\tilde\varepsilon =\varepsilon_{d}= 3.0\omega_0$. The bias range for the current suppression is larger in the case of stronger coupling to the oscillator $(g = 3.5)$.
 We obtain a hysteresis like shape for the current evolution, which is due to the multi-stability of the system. The coupling between the electronic and the mechanical system leads to a modified oscillator potential with additional minima. Switching between these states is possible and was theoretical proposed and studied by several authors \cite{Nitzan2005, Mozyrsky2006, Pistolesi2008}. 

The beginning and the ending of the hysteresis regime, where two current channels exist, are denoted by vertical dotted lines in Figure \ref{fig.:currentSD}. For the non-adiabatic case the latter regime, where two current channels exist, differs a bit from the adiabatic case.

The upper right graph of Figure \ref{fig.:currentSD} shows the current $\mathcal I_{\rm L} (x)$ for the bias value $V_{\rm{bias}}=2.0\omega_0$. Here, three fixed points occur, denoted by a cross. For $\ev{x_2^{\ast}} \approx g/2 l_0$ the effective level is situated in the middle of the transport window and following from that the current is maximal. For the two other fixed points the effective level is again situated outside the transport window and the current is small. 

By comparing the adiabatic and the non-adiabatic stationary currents, we can conclude that in the long-time limit only small differences exist. The differences are at their maximum for small times, which is clearly visible in the lower right graph of Figure \ref{fig.:currentSD}, which depicts $\mathcal I_{\rm L} (t)$. There, the oscillations in the non-adiabatic case are much larger. 

The left and right time dependent currents differ for small times $t$ in the non-adiabatic case. When we integrate $\mathcal I_{\rm L}$ and $\mathcal I_{\rm R}$ over all times the results coincide, so that there is no violation of current conservation. This is comparable to a periodically driven system with time dependent tunneling rates \cite{AlbertFlint2011}. The spectral function defined in Eq.~\eqref{eq.Spectralfunction}, is sensitive to small time differences $t-t_0$. For larger times the oscillator settles into one of the fixed points, whereas the spectral function becomes stationary and hence also the current.

\section{Two-level system}\label{sec.:twolevel}
The model we are treating in this section consists of two single dot levels which are coupled by a tunnel barrier.
Again we assume a coupling to a single bosonic mode.
The total Hamiltonian is composed of the oscillator part $\mathcal H_{\rm osc}$, cf. Eq. ~\eqref{eq.:H_osc}, 
the electronic part $ \mathcal H_{\rm e}$ and an
interaction part which describes the coupling between the oscillator and the two dots.  
In contrast to the AHM, here the oscillator couples to the difference of 
the occupation numbers with the coupling strength $\lambda$. The total Hamiltonian therefore reads
    \begin{equation}		\label{eq.:H_DQD}
	\mathcal H = \mathcal H_{\rm e} + \mathcal H_{\rm osc} 
		-\lambda \hat q (\hat d^\dag_{\rm L} \hat d_{\rm L} -\hat d^\dag_{\rm R} \hat d_{\rm R}) ,
    \end{equation}
containing the electronic part
    \begin{eqnarray}\label{eq.:H_DQD_electronic}
	   \mathcal H_{\rm e} &=& \sum_{k \alpha} \ \varepsilon_{k \alpha} c_{k \alpha}^{\dag} c_{k \alpha} \ 
		    + \ \sum_{k \alpha} \ V_{k \alpha} \ c_{k \alpha}^{\dag} d_{\alpha} \ 
		    +  V_{k \alpha}^{\ast} \  d_{\alpha}^{\dag} c_{k \alpha} \nonumber \\ & &
		    + \ \sum_{\alpha} \nu_{\alpha} d_{\alpha}^{\dag} d_{\alpha} 
		    + T_{c} \ d_{L}^{\dag} \ d_{R} + T^{\ast}_{c} \ d_{R}^{\dag} \ d_{L},
    \end{eqnarray}
where $\nu_{\alpha \in \rm{L,R}}$ denotes the left and right dot energy levels.
Here, $T_c$ denotes the tunnel coupling matrix element between the two dots.
Again, we obtain a Langevin equation, Eq.~\eqref{eq.LangevinGeneric}, with the force term 
    \begin{align} \label{forceDD}
	f\left[ x_t \right] = \lambda \ev{\sigma_z}(t) \equiv \lambda \left[\ev{n_{\rm L}}(t) - \ev{n_{\rm R}}(t)\right],
    \end{align}
where $\ev{n_{\rm L/R}}(t)$ implicitly depend on the oscillator coordinate $x_t$, cf. below.
    \begin{figure}[t]
		\centering
		\includegraphics[width=\linewidth]{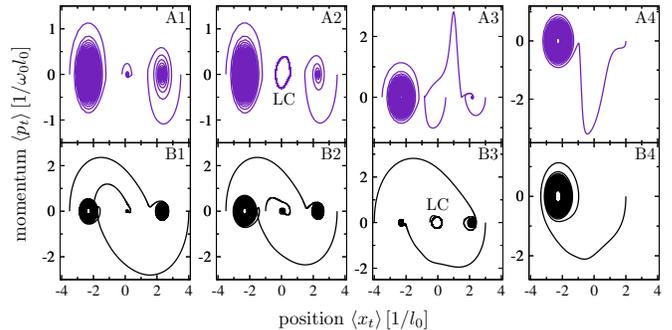}		\caption{\label{fig.:DQD_phase_space_comparison}
		Phase space portraits for various tunnel couplings $T_{c}$, increasing from left to right. 
		Upper row A: adiabatic results. Lower row B: non-adiabatic results. 
		In graphs A2 and B3 limit cycles (LC) appear.
		Explicit parameters are $\abs{T_{c}}^2 = 0.2; 0.4; 1.0; 4.0  \ \omega_{0}^2$.
		With the rate $\Gamma=2.0\omega_{0}$ and for the chemical potentials we assumed
		$\mu_{\rm L}= \omega_{0}$ and $ \mu_{\rm R}=-5 \omega_{0}$. 
		The dimensionless coupling constant is chosen as $g=2.5$, and the internal bias voltage as
		$V_{\rm int} = 5 \omega_{0}$, whereas $\nu_{\rm L}=-\nu_{\rm R}= e V_{\rm int}/2$.
		}			
    \end{figure} 
	
\subsection{Time dependent occupation}
Calculating the time dependent occupations for the two level system using Green's functions is a challenge due to the complex dependencies and couplings of the systems operators. We choose a more direct way by using the equations of motion technique, leading to a large system of coupled differential equations which have to be solved numerically.

The Heisenberg equations of motion for operators of the dots and the leads ($\tilde\nu_{\alpha}(t) = \nu_{\alpha} \mp \lambda x_t$) yield
    \begin{align}\label{eq.:DotDotHeisenberg}
         \dot{\tilde d}_{\rm L} (t) =& - i \left(\tilde\nu_{\rm L}(t) - i \frac{\Gamma}{4}\right) \tilde d_{\rm L}(t) 
                                       - i T^{\ast}_c \tilde d_{\rm R}(t)
                                       +  \sum_{k} \tilde C_{k \rm{L}}(t) ,\nonumber \\
         \dot{\tilde d}_{\rm R} (t) =& - i \left(\tilde\nu_{\rm R}(t) - i \frac{\Gamma}{4}\right) \tilde d_{\rm R}(t)
                                       - i T_c \tilde d_{\rm L}(t) 
                                       +  \sum_{k} \tilde C_{k \rm{R}}(t) ,  \nonumber \\                  
    \end{align}
where $\tilde C_{k \alpha}(t) = -i V^{\ast}_{k \alpha}  e^{-i \varepsilon_{k \alpha} t} \tilde c_{k \alpha}(0)$ and the tunneling rate equals $\Gamma_{\alpha} \equiv 2 \pi \sum_k \abs{V_{k\alpha}}^2 \delta(\omega - \varepsilon_{k\alpha})$. Again, we assume constant tunneling rates $\Gamma_{\rm L} = \Gamma_{\rm R} = \Gamma/2$.

The equations~\eqref{eq.:DotDotHeisenberg} already include the solution for the inhomogeneous differential equation for the lead operator $\tilde c_{k \alpha}$. The tilde denotes the interaction picture introduced above, cf. Eq.~\eqref{eq.TimeEvolutionOperator}. Hence, the effective time dependent energy level $\tilde\nu_{\alpha}(t)$ contains only the classical variable $x_t$.

The differential equations for the corresponding dot annihilation operators are derived in a similar manner.
Finally, one obtains an inhomogeneous system of coupled differential equations with time dependent coefficients. Multiplication with $\delta (\omega - \varepsilon_{k,\alpha})$ and summing over all $k$ states leads to
    \begin{align}\label{eq.:Sigmasystem}
              \ev{\dot{\tilde \sigma}_z}(t)      =& \ - \frac{\Gamma}{2} \ev{\tilde\sigma_z}(t)  
                                         + 2 \ \mbox{Re} \Big[  2 \ D_{\rm{RL}}(t) \nonumber\\ & \
                                         +  \int d\omega    B_{\rm{LL}}(\omega,t) 
                                         -  \int d\omega'   B_{\rm{RR}}(\omega',t)  \Big]   \nonumber\\
                \dot{D}_{RL}(t)           =&  \ i\left(\tilde\nu_{\rm R}(t)  - \tilde\nu_{\rm L}(t)  + i\frac{\Gamma}{2}\right) D_{RL}(t)
                                       - \left|T_c \right|^{2} \ev{\tilde\sigma_z}(t) \nonumber\\ & \
                                       + \int d\omega'  B_{\rm{RL}}(\omega',t) 
                                       - \int d\omega   B_{\rm{LR}}^{\dag}(\omega,t)   \nonumber\\
                \dot{B}_{\alpha, \alpha}(\omega,t) =& \
                                       - i \left(\tilde\nu_{\alpha}(t)  - \omega - i\frac{\Gamma}{4}\right) B_{\alpha, \alpha}(\omega,t) \nonumber\\ & \
                                       - B_{\alpha, \beta}(\omega,t) 
                                       +  \frac{\Gamma}{4\pi} f_{\alpha}(\omega)  \nonumber\\
                \dot{B}_{\rm{\alpha, \beta}}(\omega,t) =& \
                              -i\left(\tilde\nu_{\beta}(t)  - \omega - i\frac{\Gamma}{4}\right) B_{\alpha, \beta}(\omega,t) \nonumber\\ & \
                              +  \left|T_c \right|^{2}  B_{\alpha, \alpha}(\omega,t), \hspace{0.5cm}\alpha \neq \beta,
    \end{align}
with the definitions:
    \begin{eqnarray} \label{eq.:definitionDD}
          B_{\alpha \alpha}(\omega,t) &=& \hspace{0.6cm} i \ V_{k\alpha} \ \delta(\omega-\varepsilon_{k\alpha}) \ e^{i\varepsilon_{k\alpha} t}              
                                  \ev{\tilde c^{\dag}_{k\alpha}(0) \tilde d_{\alpha}(t)}
         \nonumber\\
         B_{\rm{RL}}(\omega,t)  &=& - T_c \ V_{k\rm R} \ \delta(\omega-\varepsilon_{k\rm R}) e^{i\varepsilon_{k\rm R} t} 
                                  \ev{\tilde c^{\dag}_{k\rm R}(0) \tilde d_{\rm L}(t)}
         \nonumber\\
         B_{\rm{LR}}(\omega,t)  &=& - T^{\ast}_c \ V_{k\rm L} \ \delta(\omega-\varepsilon_{k\rm L}) e^{i\varepsilon_{k\rm L} t} 
                                  \ev{\tilde c^{\dag}_{k\rm L}(0)\tilde d_{\rm R}(t)}
         \nonumber\\
           D_{\rm{RL}}(\omega,t)  &=&   i T_c \  \ev{\tilde d^{\dag}_{\rm R}(t)\tilde d_{\rm L}(t)}.
    \end{eqnarray}
    \begin{figure}[ht]
		\centering
		\includegraphics[width=\linewidth]{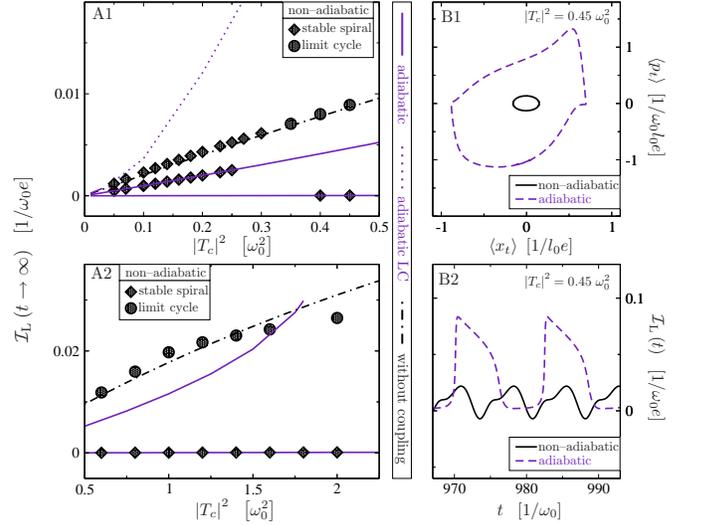}		\caption{\label{fig.:DDCurrent}
		LEFT: Current for $t \rightarrow \infty$ as a function of tunnel coupling $\abs{T_c}^2$. 
		A1 displays the results for $\abs{T_c}^2 \leq 0.5 \omega_0^2$ 
		and A2 the results in the range of $0.5 \leq \abs{T_c}^2 /\omega_0^2 \leq 2.0$.
                The symbols denote the non-adiabatic results.
		The diamonds correspond to the stable spiral situations, there the oscillation of the dynamical system disappears
		in the long-time limit and the current becomes stationary. Circles denote averaged current values for the limit cycle case, when the
	        system performs periodic oscillations. The indigo solid (dotted) lines depict the adiabatic results for the stable spiral (limit cycle)
	        case. The dashed-dotted line depicts the current without coupling. 
	        RIGHT: The graph B1 shows the phase space results for $\abs{T_c}^2 = 0.45 \omega_0^2$, here the radius for the adiabatic limit cycle is
	        much larger than in the non-adiabatic case. Below, graph B2, depicts the corresponding time dependent left current, which oscillates
	        as well.
		Explicit parameters are  $\Gamma= \omega_{0}$, $\mu_{\rm L}= \omega_{0}$ and $ \mu_{\rm R}=-5 \omega_{0}$. 
		The dimensionless coupling constant is chosen as $g=2.5$, and the internal bias voltage as
		$V_{\rm int} = 5  \omega_{0}$, whereas $\nu_{\rm L}=-\nu_{\rm R}= e V_{\rm int}/2$.
		}			
    \end{figure}
The system Eq.~\eqref{eq.:Sigmasystem} is solved numerically together with the equations of motion for the expectation values for position and momentum operator
    \begin{align}
           \dot{x}_t =& \frac{1}{m} p_t \nonumber \\
           \dot{p}_t =& - V'_{\rm{osc}}  + \lambda  \ev{\tilde \sigma_z}(t).
     \end{align}
Hence, the phase space trajectories are obtained. 

In Figure \ref{fig.:DQD_phase_space_comparison} results for $\Gamma = 2 \omega_0$ are plotted. The upper row depicts the result for the adiabatic case including the first correction term. This so-called intrinsic friction term $D\left[x_t\right]$, cf. Eq.~\eqref{eq.:LangevinGenAdiabatic}, results from the non-equilibrium electronic environment. The tunnel coupling $\abs{T_c}^2$ increases from left to right. Three fixed points appear in the range of small tunnel coupling (A1). There, the trajectories form stable spirals and run into the fixed points. 
For $\abs{T_c}^2 = 0.4 \omega_0^2$ a limit cycle appears in the middle (A2). By further increasing $\abs{T_c}^2$ the limit cycle turns into an unstable spiral(A3) and in the end only the left fixed points survives (A4). 

The second row (B) shows the results for the non-adiabatic approach.
Qualitatively the same features emerge, as the appearance of the limit cycle and the stable spirals. Comparing the adiabatic and the non-adiabatic approach, we obtain quantitative differences, like the change of the middle fixed point into a limit cycle which happens at higher values of $\abs{T_c}^2$ as in the adiabatic case. The results differ most for small times, similar to the single level case (Sec.\ref{sec.:singlelevel}). By further decreasing the tunneling rate $\Gamma$ the differences between the approaches increase. Results for a smaller tunneling rate will be presented in the next section, Sec. \ref{sec.:current}, there we calculate the electronic current for $\Gamma = \omega_0$.

We also mention that the appearance of the limit cycle in this system is possible due to energy transfer processes between the electrons and the oscillator. In the stable spiral case, the influence of the electrons leads to damping of the oscillator. For example for $\tilde \nu_{\rm L} <  \tilde \nu_{\rm R}$,  the electrons need energy to pass through the system.

The requirements for a damped dynamical system to exhibit limit cycles is the additional appearance of positive friction. For our system this means, that energy transfer processes occur which lead to the acceleration of the oscillator. 
The occurrence of positive friction is possible when  $\tilde \nu_{\rm L} >  \tilde \nu_{\rm R}$ and the electrons can transfer energy to the oscillator, cf.\cite{Hussein2010}.

\subsection{Current} \label{sec.:current}
The current through lead $\alpha$ is derived via the Heisenberg equations of motion and yields
    \begin{align}
           \mathcal I_{\alpha} (t) = - e 
                        \left[\Gamma_{\alpha}  N_{\alpha} \left[ x_{t}\right] 
                            - \int d\omega \mbox{Re} \left[ B_{\alpha \alpha} (\omega, t)\right] \right].
    \end{align}             
The left graphs of Figure \ref{fig.:DDCurrent} depict the stationary current $\mathcal I_{\rm L} (t \rightarrow \infty)$ as a function of the tunnel coupling $\abs{T_c}^2$. In the adiabatic case and for small values of $\abs{T_c}^2$ (A1) we observe two fixed points and one limit cycle leading to a tri-stable current. In the limit cycle case the current oscillates in time. The corresponding averaged current (dotted line) is not completely shown in (A1), due to the large values, since the current increases further until $\abs{T_c}^2 = 0.48 \omega_0^2$. There, the limit cycle disappears and two fixed points remain until $\abs{T_c}^2=1.8 \omega_0^2$ (A2). We also obtain two fixed points in the small range of $\abs{T_c}^2 \leq 0.03 \omega_0^2$, which is not dissolved in graph A1 of Figure \ref{fig.:DDCurrent}.

The current corresponding to the fixed point $x_{\ast} \simeq 2 / l_0$ (solid line below $g=0$ case) increases approximately in the same fashion as in the case without coupling. In this regime the left effective level lays inside the transport window ($\widetilde \nu_{\rm{L,R}} \simeq \mp 2.5 \omega_0$). 
For large tunnel coupling one fixed point persists, $x_{\ast} \simeq -2.3 / l_0$, and the corresponding current (lowest solid line) is strongly suppressed compared to the case without coupling. There, both effective levels $\widetilde \nu_{\rm{L,R}} \simeq \pm 7.5 \omega_0$ are clearly situated outside the transport window. Therefore tunneling through the two level system is rarely possible. 

In the left graphs of Figure \ref{fig.:DDCurrent} the symbols denote the non-adiabatic current in the long-time limit. Here, the system has also two fixed points, but the limit cycle range is much larger, $0.35 \leq \abs{T_c}^2/\omega_0^2 \leq 6.5$. For $\abs{T_c}^2 \leq 0.35 \omega_0^2$ we observe two stable fixed points and by increasing the tunnel coupling the middle stable spiral turns into a limit cycle and the mechanical system performs periodic oscillations. For the latter case, the circles in Figure \ref{fig.:DDCurrent} denote the averaged current.

The non-adiabatic current corresponding to the middle fixed point/limit cycle follows the result without coupling. As long as the fixed point $x^{\ast} \simeq 0.03 / l_0$ is stable the resulting effective level is approximately $\widetilde \nu_{\rm{L,R}} \simeq \nu_{\rm{L,R}}$ as in the case without coupling. In graph B1 of Figure \ref{fig.:DDCurrent}, the phase space trajectories are plotted for the case when the system performs periodic oscillations. The limit cycle, corresponding to the non-adiabatic results, runs in small cycles about the origin and following from that, the averaged current is similar to the case without coupling. In the adiabatic case the radius is much larger and the shape of the limit cycle is not smoothly circular. Hence, the current is much larger then in the non-adiabatic case (A1). 
In graph B2 the related time dependent current is depicted. Frequency and amplitude differ strongly in both cases.
    \begin{figure}[h]
		\centering
		\includegraphics[width=\linewidth]{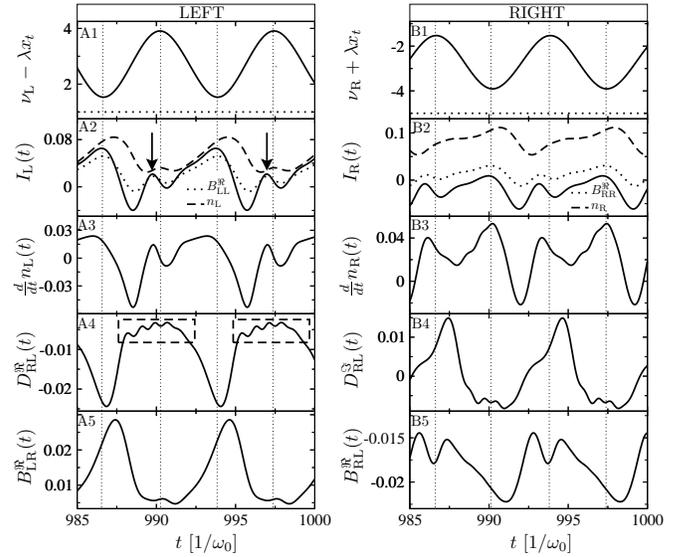}		\caption{\label{fig.:AnalyseFunctionsDD}
		Time evolution for the current and the corresponding correlation functions for $\abs{T_c}^2 = \omega_0$ in the
		non-adiabatic limit cycle case. The vertical dotted lines correspond to the maxima/minima of the effective level 
		 $\tilde\nu_{\alpha} = \nu_{\alpha} \mp g x_{t}/l_0$.
		The first row depicts the position for the effective left (A1) and right (B1) level and
		the dotted line corresponds to the chemical potentials $\mu_{\rm L}= \omega_{0}$ and $ \mu_{\rm R}=-5  \omega_{0}$. 
		The results for the left/right current are plotted in graph A2/B2, together with the occupation for the left/right level and the real part
		of the correlation function $B_{\rm{LL/RR}}$. The time derivative of the level occupation is depicted in row three. 
		Row 4 shows the result for the real (A4) and the imaginary (B4) part of the dot-dot-correlation function $D_{\rm {RL}}$.
	        The real parts of $B_{\rm{LR/RL}}$ are depicted in row 5.
		The dimensionless coupling constant is chosen as $g=2.5$, and the internal bias voltage as
		$V_{\rm int} = 5  \omega_{0}$, whereas $\nu_{\rm L}=-\nu_{\rm R}= e V_{\rm int}/2$.
		}			
    \end{figure}
The frequency of the current oscillations is equal to the oscillator frequency (non-adiabatic: $\omega \approx 0.86 \omega_0 $). 

The current reaches its maximum when the distance between the left (right) effective level and the left (right) chemical potential is minimal (maximal). This is clearly visible in Figure \ref{fig.:AnalyseFunctionsDD}, where the time evolution for current and the different correlation functions defined above, cf. Eq.~\eqref{eq.:definitionDD}, are depicted. 
The first row shows the behavior of the oscillating effective levels, the value of the chemical potential is also plotted in these graphs (dotted line).
In the first graph of the second row the current for the left lead is maximal when the effective level is minimal as mentioned above. The current decreases when the left level increases its distance to the transport window. 
	
The additional current peak (arrows in A2 of Fig. \ref{fig.:AnalyseFunctionsDD}) near the maximum of the left effective level does not appear in an adiabatic approach, where the current follows the position of the levels. This peak is related to internal coherent electronic oscillations between the two dots.
These oscillations are visible in the real part of the $D_{RL}$ function depicted in graph A4 of Figure \ref{fig.:AnalyseFunctionsDD} (denoted by dashed boxes), with frequencies that match the time dependent Rabi frequency $\omega_{R}(t) = \sqrt{\tilde\nu_{\rm L}(t) - \tilde\nu_{\rm R}(t) + 4 \abs{T_c}^2}$.

In the adiabatic case, the time-resolved current for the right lead is equal to the current through the left lead with opposite sign.
For the non-adiabatic case, right and left time-resolved currents are different, but their time-averages coincide. If we are in the long-time limit, and the system performs no oscillations, left and right current are equal.
In contrast, in the limit cycle case we obtain a driven system leading to currents $\mathcal I_{\rm{L,R}}$ whose time dependence differ, since charge temporarily accumulates in the dots.
	
\section{Conclusion}
By comparing the adiabatic and non-adiabatic results for the single-level system we obtain a good qualitative agreement. In principle, the same features arise, as bistability and a hysteresis-like $\mathcal I-V $ characteristic are observed in both cases. The largest deviations are observed for small times, but in the long time limit the results predominantly coincide.

For the two-level case the differences are much larger. Qualitatively we observe similar properties, but the quantitative predictions of the adiabatic approach do not match the results for the non-adiabatic system where the oscillator and the electrons act on the same timescale. The electron-oscillator interaction leads to multiple current channels like in the single-level system. Additionally, we observe limit cycles of the dynamical system leading to periodic oscillations of the current. In this regime, the system acts as a DC-AC-transformer.	

\textbf{Acknowledgments.} This work was supported by projects DFG BR 1528/7-1, DFG BR 1528/8-1 and the Rosa Luxemburg foundation.

\end{document}